\documentclass[twocolumn,twoside]{revtex4}
\usepackage{graphicx}
\usepackage{fancyhdr}
\begin{document}

\title{Prospects for $\tau$ Lepton Physics at Belle II }

%

\author{D. Rodr\'iguez P\'erez\\
on behalf of Belle II Collaboration}
\affiliation{Facultad de Inform\'atica Culiac\'an, Universidad Aut\'onoma de Sinaloa,\\
Cd. Universitaria, CP 80013, Sinaloa, M\'exico.}

\begin{abstract}
The Belle II experiment is a substantial upgrade of the Belle detector and will operate at  the SuperKEKB energy-asymmetric $e^+e^-$ collider. The design luminosity of the machine is 8$\times10^{35}$ cm$^{-2}$s$^{-1}$ and the Belle II experiment aims to record 50 ab$^{-1}$ of data, a factor of 50 more than its predecessor. From February to July 2018, the machine has completed a commissioning run and main operation of SuperKEKB has started in March 2019. Belle II has a broad $\tau$ physics program, in particular in searches for lepton flavor and lepton number violations (LFV and LNV), benefiting from the large cross section of the pairwise $\tau$ lepton production in $e^+e^-$ collisions. We expect that after 5 years of data taking, Belle II will be able to reduce the upper limits on LFV and LNV $\tau$ decays by an order of magnitude. Any experimental observation of LFV or LNV in $\tau$ decays constitutes an unambiguous sign of physics beyond the Standard Model, offering the opportunity to probe the underlying New Physics. In this talk we will review the $\tau$ lepton physics program of Belle II.
\end{abstract}

\maketitle

\thispagestyle{fancy}

\section{Introduction} 
The SuperKEKB \cite{Ohnishi2013} accelerator is upgraded from KEKB and its target luminosity is $8\times 10^{35}$ cm$^{-2}$s$^{-1}$,  40 times higher than KEKB. SuperKEKB collides electrons and positrons at the $\Upsilon(4S)$ resonance energy, producing a large amount of B meson pairs, a B-factory. However, the cross section of the process $e^+e^- \to \tau^+\tau^-$ at the $\Upsilon(4S)$ resonance energy is of the same order as the production of a B pair, then, SuperKEKB is also a $\tau$ lepton factory.

The Belle II \cite{Abe2010} detector has been upgraded from Belle detector to cope with higher luminosity and higher expected beam backgrounds. It is a second generation B-factory and offers unique capabilities to study $B$ and $\tau$ decays with neutrinos in the final states, with respect to experiments at hadron colliders.

The Belle II program is based on the success of the BaBar at SLAC and Belle at KEK, the first generation of B-factories, which discovered and established violation of charge-parity symmetry (CPV) in $B$ meson dynamics, furthermore they performed precision measurements of $\tau$ properties, such as the mass, lifetime and branching fractions of leptonic and semileptonic decays. Additionally, they imposed limits in electric dipole moment, lepton flavor violation (LFV) and lepton number violation (LNV) decays \cite{Bevan2014}.

Now, Belle II has an ambitions program in $\tau$ physics, aiming to move down the upper limit of the rate of LFV and LNV $\tau$ decays by two order of magnitude.

\section{The Belle II Experiment}
The Belle II experiment is a detector coupled to the SuperKEKB accelerator, located in Tsukuba, Japan. Belle II keeps the design of the previous detector Belle, with major upgrades in each of their subsystems. The main modifications are:

\begin{itemize}
  \item The vertex detector, which contains two layers of DEPFET pixel (PXD) and four layers of silicon strips (SVD), improving the resolution respect to Belle.
  \item The central drift chamber (CDC) has a larger volume with smaller drift cells.
  \item A completely new particle identification system, using aerogel ring-imaging Cherenkov detectors.
  \item Faster electronics in general. 
\end{itemize}

A complete description of the Belle II detector can be read at reference \cite{Abe2010}.

From April to July of 2018, Belle II performed the Phase II of commissioning. Detector recorded 500 pb$^{-1}$ of data at $\Upsilon(4S)$ energy with the BEAST II detector installed, instead of the vertex detector. The BEAST II contains slice of the vertex detector and radiation monitors used in beam background studies \cite{Lewis2019}.
Later in 2018, BEAST II was replaced by the vertex detector and in March 2019, Belle II was started Phase III of data with all the subsystems installed, expecting a full dataset of 50 ab$^{-1}$ by the end of the data taking, in 2025.
 
\section{First Results of $\tau$ Lepton Physics at Belle II}

\subsection{Reconstruction of $\tau$ Pair Production}
The reconstruction of the tau pair production $e^+e^- \to \tau^+\tau^-$ is performed searching 3-1 prong events in a data sample of 291 pb$^{-1}$. Events in data are required to fire the CDC trigger. Only four charged tracks per event are accepted, with zero net charge and splitting the decay products into two opposite hemispheres by a plane perpendicular to the thrust axis $\vec{\bar{n}}_{thr}$, defined such that 

\begin{equation}
V_{thr} = \frac{\sum_i \vert \vec{p}_i^{\,cm} \cdot\hat{n}_{thr} \vert}{\sum \vert \vec{p}_i^{\,cm} \vert}
\label{thrust}
\end{equation}
is maximized, with $\vec{p}^{\,cm}_i$ being the momentum in the center-of-mass system (CMS) of each charged particle and photon. Signal side hemisphere is defined as the one containing a 3-prong decay, while the tag side should contain the 1-prong decay. A pion mass hypotesis is used for all charged tracks, looking for $\tau \to 3\pi\nu$ events in the signal side.

\begin{figure}[h]
\centering
\includegraphics[width=80mm]{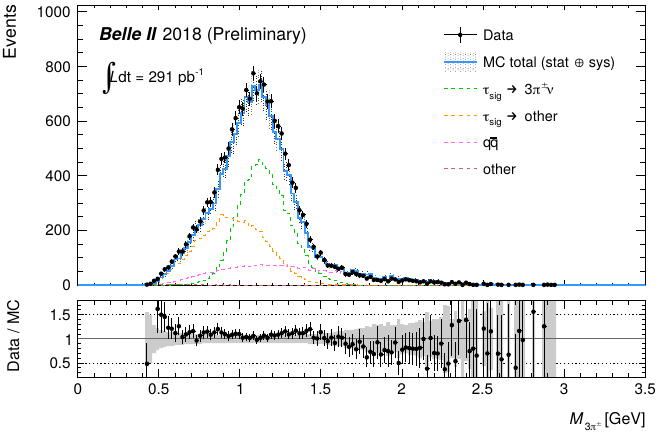}
\caption{Invariant mass distribution of the three pions coming from $\tau \to 3\pi\nu$ candidates reconstructed in Phase II data  \cite{Michel2018}. Events in data are required to fire the CDC trigger. MC is rescaled to a luminosity of 291 pb$^{-1}$ and reweighted according to the trigger efficiency measured in data. The error band on the total MC includes the MC statistical uncertainty, the luminosity uncertainty and the uncertainty associated with the trigger efficiency reweighting.} \label{invmass}
\end{figure}

\begin{figure}[h]
\centering
\includegraphics[width=80mm]{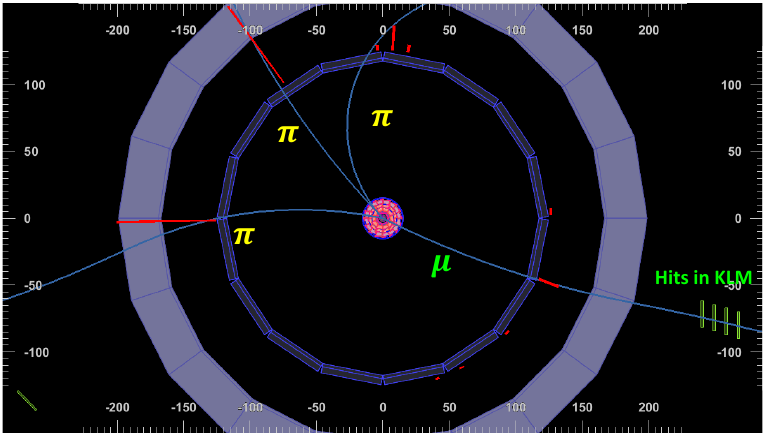}
\caption{Event desplay of the Belle II detector showing a 3-1 prong event, likely a $e^+e^- \to (\tau_{sig} \to 3\pi\nu_\tau)(\tau_{tag} \to \mu\nu_\tau\bar{\nu}_\mu)$ candidate reconstructed in Phase II data \cite{Michel2018}.} \label{candidate}
\end{figure}

After further selection criteria are applied, 9800 events remain as $\tau$ pair candidates. Figure \ref{invmass} shows the invariant mass distribution of the three charged pions coming from $\tau\to3\pi\nu$ candidates, with Monte Carlo (MC) simulated events superimposed. Figure \ref{candidate} presents the event display of the Belle II detector with one of the candidates who pass the cuts.

\subsection{$\tau$ Lepton Mass Measurement}
A first $\tau$ lepton mass measurement at Belle II is performed following the method develop by the ARGUS collaboration \cite{Albrecht1992}. The pseudomass $M_{min}$, defined by 

\begin{equation}
  M_{min} = \sqrt{M^2_{3\pi}+2(E_{beam}-E_{3\pi})(E_{3\pi}-P_{3\pi})},
  \label{minM}
\end{equation}
is obtained for each $\tau\to3\pi\nu$ candidate. Here, $E_{beam}$ is the energy of one of the beams in CMS and $M_{3\pi}$, $E_{3\pi}$ and $P_{3\pi}$ represent the invariant mass, the energy and the momentum of the hadronic system of the three pions in CMS, respectively.

An empirical probability density function (p.d.f.) is used to estimate the $\tau$ lepton mass. The edge p.d.f. used is described by

\begin{eqnarray}
F(M_{min};a,b,c,m^*) & = & (a*M_{min}+b)\times \nonumber\\
  &   & arctan[(m^*-M_{min})/c] + \nonumber \\
  &   & P_1(M_{min}),
\label{pdf}
\end{eqnarray}
in which $a$, $b$ and $c$ are real values and the paramenter $m^*$ is an estimator of the $\tau$ lepton mass.

\begin{figure}[h]
\centering
\includegraphics[width=80mm]{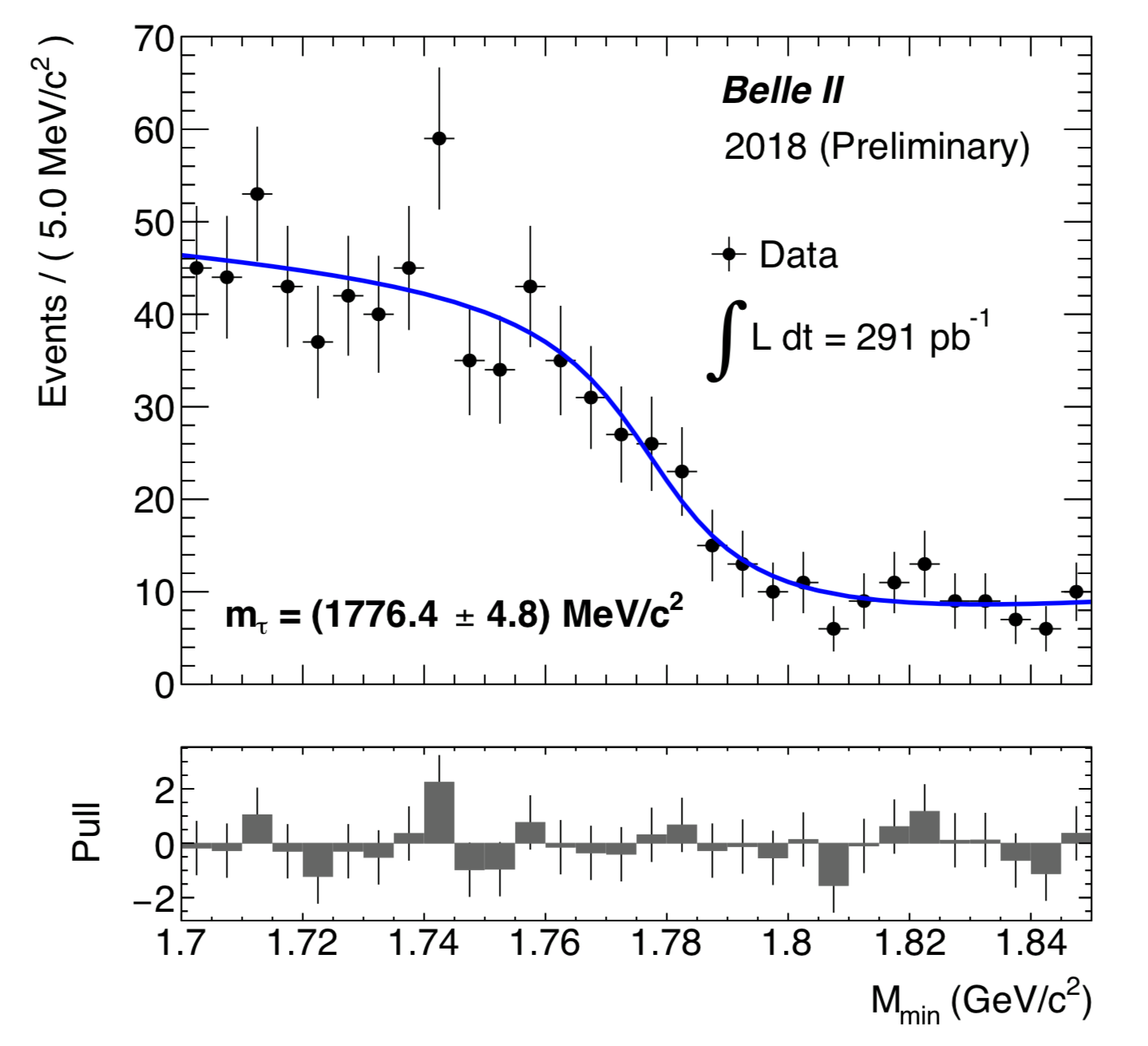}
\caption{Distribution of pseudomass $M_{min} = \sqrt{M^2_{3\pi}+2(E_{beam}-E_{3\pi})(E_{3\pi}-P_{3\pi})}$ of $\tau^-_{sig} \to 3\pi\nu$ candidates reconstructed in Phase II data. The blue line is the result of an unbinned maximum likelihood fit, using an edge function $(a*M_{min}+b)\cdot\arctan [(m^*-M_{min})/c] + P_1(M_{min})$, in which $m^*$ estimates the $\tau$ lepton mass. A mass of $m_\tau = (1776.4\pm 4.8$(stat)) MeV/c$^2$ is measured \cite{Michel2018}.} \label{taumass}
\end{figure}

A fit of the p.d.f. (\ref{pdf}) in the pseudomass region from 1.70 to 1.85 GeV/c$^2$, yields a mass measurement of $m_\tau = (1776.4\pm4.8$(stat)) MeV/c$^2$. Figure \ref{taumass} shows the pseudomass distribution of the $\tau\to3\pi\nu$ candidates, with the p.d.f fitted superimposed. The result is in good agreement with the measurements from the previous experiments \cite{PDG2018}.

\section{Prospects for $\tau$ Lepton Physics}
The goal of Belle II experiment is to focus on precision measurements and extrapolation of new physics (NP) in rare physics processes, and with 45 billions of $e^+e^- \to \tau^+\tau^-$ events expected at the end of the data taking,  the study of $\tau$ physics will be possible. Prospect for $\tau$ lepton physics at Belle II are briefly discribed . Further details and a more complete description may be found in the Belle II Physics Book \cite{Kou2018}.
 
\subsection{Lepton Flavor Violation in $\tau$ Decays}
The discovery of neutrino oscillations has demonstrated that lepton flavor number are not conserved in the neutrino sector and claims for an extended Standard Model (SM) with massive neutrinos. But, if the SM is extended to include neutrino masses only, the branching ratio of lepton flavor violation (LFV) processes is too small, $\sim10^{-54}$, to be observed \cite{Petcov1997,Hernandez2019}. Then, Belle II will be very competitive to explored NP in this field for its relatively low background \cite{Celis2014}.

The golden channels for studying charged LFV are $\tau\to3\mu$ and $\tau\to\mu\gamma$. The first one is a purely leptonic state and the background is suppressed; the second one has the largest LFV branching fraction in models where the decay is induced by one-loop diagrams with heavy particles \cite{Hisano2002,Arganda2006}.

\begin{figure}[h]
\centering
\includegraphics[width=80mm]{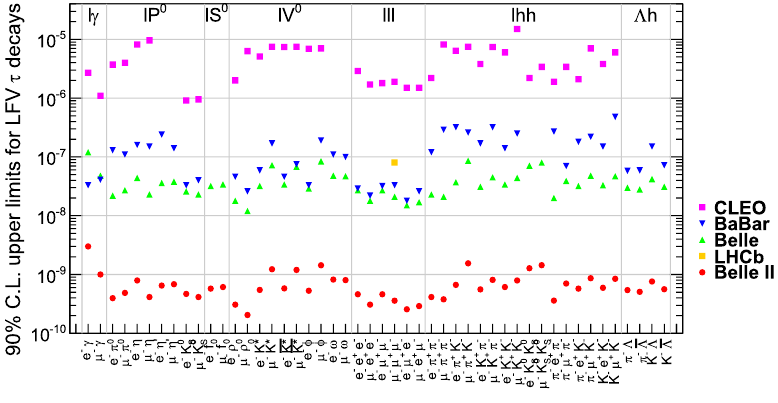}
\caption{Current 90$\%$ C.L. upper limits for the branching fraction of LFV $\tau$ decays. Limits imposed by CLEO, BaBar and LHCb are showed. Additionally, prospects for limits to be imposed by Belle II are indicated with red circles, assuming an integrated luminosity of 50 ab$^{-1}$ \cite{Kou2018}.} \label{branchings}
\end{figure}

Figure \ref{branchings} shows the prospects for upper limits to be imposed in $\tau$ LFV decays according to sensitivity studies described at \cite{Kou2018} and, for comparison, the limits imposed for previous experiments. With the full dataset expected for the Belle II experiment, 50 ab$^{-1}$, the upper limit for the branching fraction of LFV decays $\tau$ will be reduced by two orders of magnitude.

\subsection{CP Violation in $\tau$ Decays}
It is also expected that Belle II will measure CP asymmetries in $\tau$ decays at a level that bounds many models of NP in a complementary way to the LFV searches. For example, CPV in $\tau \to K^0_S \pi\nu$. This decay of the $\tau$ lepton to final states containing a $K^0_S$ meson will have a nonzero decay-rate asymmetry $A_\tau$, defined by

\begin{equation}
  A_\tau = \frac{\Gamma(\tau^+\to\pi^+K^0_S\bar{\nu}_\tau)-\Gamma(\tau^-\to\pi^-K^0_S\nu_\tau)}{\Gamma(\tau^+\to\pi^+K^0_S\bar{\nu}_\tau)+\Gamma(\tau^-\to\pi^-K^0_S\bar{\nu}_\tau)},
  \label{asyT}
\end{equation}
due to CP violation in the kaon sector. The SM prediction \cite{Grossman2012,Bigi2005} yields 

\begin{equation}
  A^{SM}_\tau = (3.6\pm0.1)\times10^{-3}.
  \label{asyTSM}
\end{equation}

On the experimental side, BaBar is the only experiment that has measured $A_\tau$ \cite{Lees2012}, getting 

\begin{equation}
  A^{BaBar}_\tau = (-3.6\pm2.3\pm1.1)\times10^{-3},
  \label{asyTBB}
\end{equation}
which is 2.8$\sigma$ away from the SM prediction (\ref{asyTSM}). An improved measurement of $A_\tau$ is a priority at Belle II.

CP violation could alse arise from a charged scalar boson exchange. It can be detected as a difference in the decay angular distributions. Belle searched for CP violation in angular observables of the decay $\tau \to K^0_S \pi \nu$ \cite{Bischofberger2011}, in which almost all contributions to systematic uncertainty depend on the control sample statistics. So, it is expected that the uncertainties at Belle II will be improved by a factor of $\sqrt{70}$, given the integrated luminosity projected.

\section{Conclusions}
Belle II has recorded successfully $\sim500$ pb$^{-1}$ of data during the first collisions performed at SuperKEKB during Phase II. The data has been used mostly for beam background and detector performance studies, showing a healthy operation of all the subsystems. This year, full physics program has started and by the end of the experiment, in 2025, Belle II is expected to collect 50 ab$^{-1}$ of data.

The preliminary result on $\tau$ lepton mass measurement obtained from Phase II data, $m_\tau = (1776.4 \pm 4.8)$ MeV, is in good agreement with the measurements reported by previous experiments and the average $\tau$ mass value by the PDG. Systematic uncertainties were not consider. Additionally, a good agreement between data and simulation is observed.

The $\tau$ lepton physics program at Belle II will take advantage of the largely integrated luminosity expected, allowing the study of several topics. Limits in branching ratio of LFV decays, CP violation asymmetry parameters will be improved by two orders of magnitude, but a careful analysis of systematic uncertainties is required.

\bigskip 

\end{document}